\documentclass[a4paper,12pt, twoside]{article}
\usepackage{amsmath,amsthm,amssymb,graphics,graphicx,color,tabularx,cases,longtable}
\usepackage{setspace}
\usepackage[numbers,sort&compress]{natbib}
\begin{document}

\title{General considerations of the electrostatic boundary conditions in oxide heterostructures}

\author{Takuya Higuchi $^{1}$ and Harold Y. Hwang $^{2, 3}$ \\
\footnotesize {\it $^1$ Department of Applied Physics, University of Tokyo, Hongo, Tokyo 113-8656, Japan} \\
\footnotesize {\it $^2$ Department of Applied Physics and Stanford Institute for Materials and Energy Science, }\\ 
\footnotesize {\it Stanford University, Stanford, California 94305, USA}\\
\footnotesize {\it $^3$ Correlated Electron Research Group (CERG), RIKEN-ASI, Saitama 351-0198, Japan} \\
}
\date{}

\maketitle

\section{Introduction}

\begin{figure}[]
\begin{center}
\includegraphics[width=10cm]{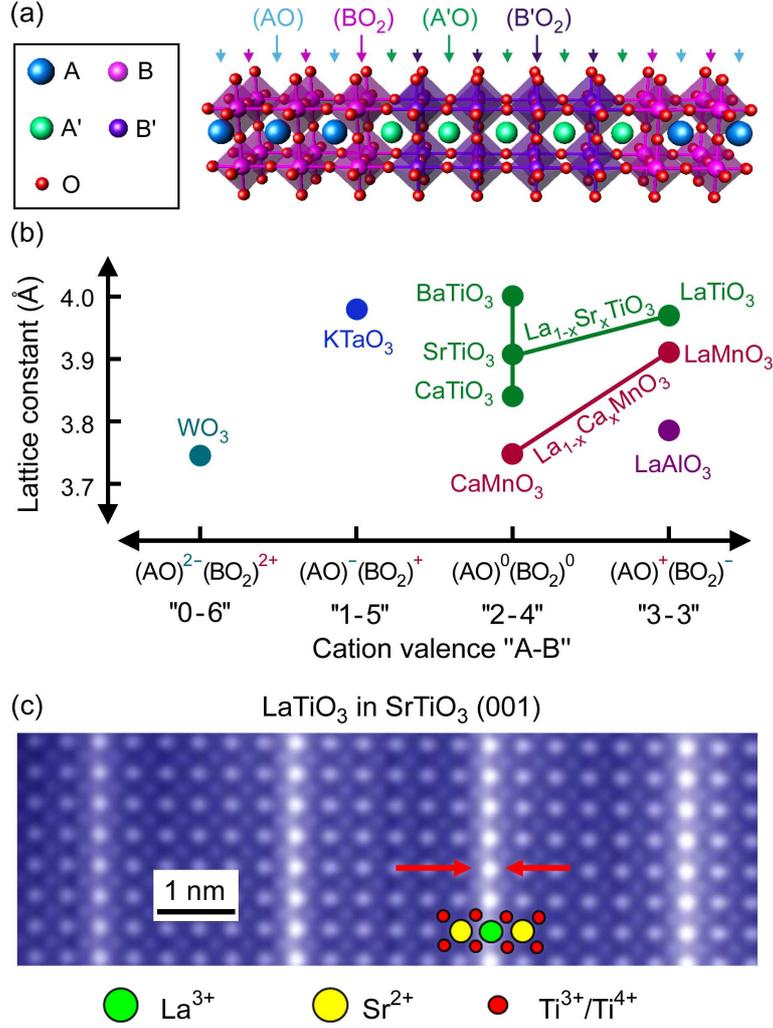}
\caption{\label{Perovskite}
(a) Schematic illustration of ideal heterointerfaces of two perovskites ABO$_3$ and A$^\prime$B$^\prime$O$_3$ stacked in the [001] direction. 
(b) Charge sequences of the AO and BO$_2$ planes of perovskites plotted together with their pseudocubic lattice parameters.
(c) Scanning transmission electron microscopy image of a LaTiO$_3$/SrTiO$_3$ (001) superlattice 
(Ohtomo {\it et al.} \cite{Ohtomo2002}).
}
\end{center}
\end{figure}

When the size of materials is comparable to the characteristic length scale of their physical properties,
novel functionalities can emerge.
For semiconductors, this is exemplified by the ``superlattice''
concept of Esaki and Tsu, 
where the width of the repeated stacking of different semiconductors 
is comparable to the ``size'' of the electrons,
resulting in novel confined states 
now routinely used in opto-electronics \cite{Esaki1970}.
For metals, a good example is magnetic/non-magnetic
multilayer films that are thinner than the spin-scattering length, from which
giant magnetoresistance (GMR) emerged \cite{Baibich1988, Binasch1989},
used in the read heads of hard disk drives.
For transition metal oxides, a similar research program is currently underway, 
broadly motivated by the vast array of physical properties that they host.  
This long-standing notion has been recently invigorated by the development of atomic-scale growth and probe techniques, 
which enables the study of complex oxide heterostructures approaching the precision idealized in Fig.~\ref{Perovskite}(a).  
Taking the subset of oxides derived from the perovskite crystal structure, the close lattice match across many transition metal oxides presents the opportunity, in principle, to develop a ``universal'' heteroepitaxial materials system.

Hand-in-hand with the continual improvements in materials control, 
an increasingly relevant challenge is to understand the consequences of the electrostatic boundary conditions which arise in these structures.  
The essence of this issue can be seen in Fig.~\ref{Perovskite}(b), 
where the charge sequence of the sublayer ``stacks'' for various representative perovskites is shown in the ionic limit, in the (001) direction.  
To truly ``universally'' incorporate different properties using different materials components, be it magnetism, ferroelectricity, superconductivity, etc., 
it is necessary to access and join different charge sequences, 
labelled here in analogy to the designations ``group IV, III-V, II-VI'' for semiconductors.  As we will review, interfaces between different families creates a host of electrostatic issues.  They can be somewhat avoided if, as in many semiconductor heterostructures, only one family is used, 
with small perturbations (such as n-type or p-type doping) around them\footnote{These effects can in principle also be reduced by choosing a (110) growth orientation, but other aspects of stability may be limiting \cite{Mukunoki}.}. 
However, for most transition metal oxides, this is greatly restrictive.  
For example, LaMnO$_3$ and SrMnO$_3$ are both insulators in part due to strong electron correlations, 
and only in their solid solution does ``colossal magnetoresistance'' emerge in bulk \cite{Tokura1994}.  
Similarly, the metallic superlattice shown in Fig.~\ref{Perovskite}(c) can be considered a nanoscale deconstruction of (La,Sr)TiO$_3$ to the insulating parent compounds. 
Therefore the aspiration to arbitrarily mix and match perovskite components requires a basic understanding of, and ultimately control over, these issues. 

In this context, here we present basic electrostatic features that arise in oxide heterostructures which vary the ionic charge stacking sequence.  
In close relation to the analysis of the stability of polar surfaces and semiconductor heterointerfaces, 
the variation of the dipole moment across a heterointerface plays a key role in determining its stability. 
Different self-consistent assignments of  the unit cell are presented, allowing the {\it polar discontinuity} picture to be recast in terms of an equivalent {\it local charge neutrality} picture.  
The latter is helpful in providing a common framework with which to discuss electronic reconstructions, local-bonding considerations, crystalline defects, and lattice polarization on an equal footing, all of which are the subject of extensive current investigation.

\section{The \textit{polar discontinuity} picture}

\subsection{Stability of ionic crystal surfaces}

\begin{figure}[]
\begin{center}
\includegraphics[width = 4cm]{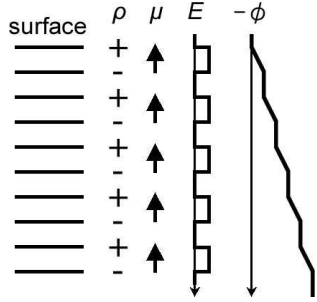}
\caption{\label{local-div} 
Schematic illustration of the charge density $\rho$, 
the dipole $\mu$ in the unit cell starting from the top-most layer,
the electric field $E$ induced by the dipoles,
and the electrostatic potential $\phi$
at the surface of an ionic crystal with dipole moment in each unit cell.}
\end{center}
\end{figure}

The surface of crystals determines many of their physical, mechanical and chemical properties.
Due to the lack of translational symmetry in the perpendicular direction, 
the stable charge distribution at the surface can be completely different from that of the bulk,
and the surface may reconstruct in a manner different from the bulk states.
Imagine an ideal ionic crystal which consists of charged ions bound together by their attractive interactions,
and all the ions are taken as fixed point charges.
Since the charges are locally preassigned to the ions in this model,
the ideal ionic surface apparently requires no reassignment of the charges from that of the bulk.
However, the electrostatic potential in an ionic crystal diverges
when there is a dipole moment in the unit cell perpendicular to the surface.
The potential\footnote{Note that literature on this topic uses both the electrostatic potential (for a positive test charge) and the electron potential energy (as in band diagrams) --- we use the former here.} $\phi$ should be constant in vacuum 
in the absence of external fields, and the potential can be obtained by integrating
the electric field caused by the charged sheets, as shown in Fig.~\ref{local-div}.
A finite shift in the potential emerges due to the dipole moment of each unit cell,
and as the unit cells are stacked, so the potential grows, and diverges into the crystal.
Due to this effectively infinite surface energy, such surfaces cannot exist without reconstructions, 
and the stability of an ionic surface randomly cut from the bulk
is not trivial without knowing the stacking sequence of the charged sheets precisely.
This surface instability and the associated reconstructions have been indeed observed by means of low-energy electron diffraction (LEED)
and ion scattering,
where absorption of foreign atoms,  surface roughening, or changes in surface stoichiometry were found 
\cite{Kinniburgh1975, Netzer1975, Benson1967}.

\begin{figure}[]
\begin{center}
\includegraphics[width = 11cm]{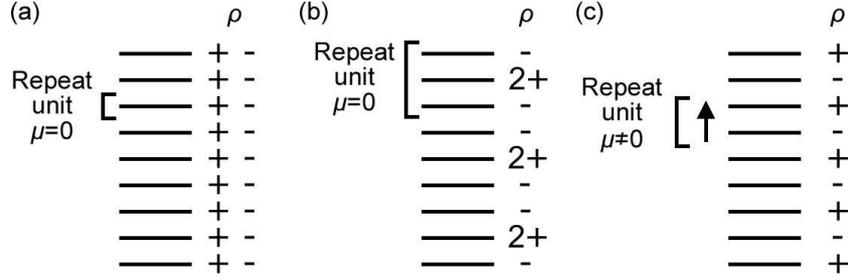}
\caption{\label{local-Tasker} 
Distribution of charges $\rho$ on planes for three stacking sequences parallel to the surface.
(a) Type 1, (b) type 2, and (c) type 3 (Tasker \cite{Tasker1979}).}
\end{center}
\end{figure}

In order to survey the stability of such surfaces, Tasker introduced a classification of the surfaces
from the viewpoint of the charge of the atomic sheets and the dipole 
in a unit cell which \textit{starts from the top-most layer} \cite{Tasker1979}.
Note he discussed the stability of \textit{bulk frozen} surfaces\footnote{
This definition of ``\textit{bulk frozen}'' follows description of Goniakowski \textit{et al.} \cite{Goniakowski2008}.}, where the top-most layer is one of the 
constituent atomic sheets of the bulk crystal and has no reconstruction.
Tasker described three types of the surfaces, as shown in Fig.~\ref{local-Tasker}:
\begin{itemize}
\item Type 1 has equal numbers of anions and cations on each plane, and therefore 
the unit cell has no dipole moment. 
For example, the (001) and (110) surfaces of the rocksalt structure $MX$ (e.g. NaCl, MgO, NiO) 
are classified as this type.
\item Type 2 has charged planes, but no net dipole moment perpendicular to the surface.
The anion $X$ terminated (001) surface of the fluorite structure $MX_2$ (e.g. UO$_2$, ThO$_2$) 
is an example.
\item Type 3 has charged planes and a net dipole moment normal to the surface.
Examples include the (111) surface of the rocksalt structure, and (001) or (111) surfaces of the zincblende structure $MX$ (e.g. GaAs, ZnS).
\end{itemize}
In Tasker's classification, type 1 and type 2 surfaces are stable while type 3 is not,
since the instability of the surface comes from the stacking of the dipole in each unit cell.
Macroscopically, the instability of the type 3 surface
arises from the change in the potential slope when crossing the surface.
The term ``polar surface'' can be defined following this classification,
namely we call a surface ``non-polar'' when it is  type 1 or type 2,
and ``polar'' when it is type 3.

These definitions are analogous to the definition of polar crystals,
although polar surfaces and surfaces of polar crystals are not equivalent.
Dielectric polarization is observed when an electric field is applied to a material,
but even in the absence of the field, some crystals retain a ``spontaneous'' polarization \cite{Mermin}.
Only 10 out of 32 point groups show this behavior, and their members are called polar crystals,
while the others are non-polar.
Twenty-one out of 32 point groups do not have inversion centers, and the polar crystals
are included among them.
When a material shows macroscopic spontaneous polarization,
the electrostatic potential of one end is different from the other end
as a result of the stacking of the dipole in each unit cell,
and the surface usually has ``compensating'' charge to reconcile this potential difference.
Since Tasker took unit cells from the top-most layer,
even crystals with inversion symmetry can show dipoles in the unit cells in his model.
For example, although NaCl is cubic and has inversion symmetry,
its (111) surface is classified into type 3.
Therefore, the word ``polar'' should be used with some care
since it has different meanings in different contexts.


\subsection{Stability of covalent surfaces}

At the surface of covalent crystals, lacking full coordination, the top-most atoms have valence electrons which are not used for bond formation. 
These non-bonding electrons are called dangling bonds, 
and have higher energy than the bonding electrons,
which causes the movement of the atom positions to decrease their number \cite{Harrison1976}.

In a covalent crystal, since the bonds are formed as a hybridization of the valence electrons of charge-neutral atoms,
one can describe the charge distribution
starting from a bulk unit cell which is charge neutral and dipole-free.
However, when the electronegativity of the atoms are different e.g., Ga and As in GaAs, 
charge transfer between anions and cations occurs, similar to the ionic case.
This charge transfer is realized by the difference of the contribution of each bond,
namely $1+\alpha$ electrons to the anions and  $1-\alpha$ electrons to the cations.
Here $\alpha$ is a parameter to describe the ionicity of the bond,
determined by the electronegativity of the two bonding atoms.
As a consequence, the unit cell can have a dipole moment normal to the surface,
which causes the same instability as that in the ionic crystal case.
Therefore, even in a covalent crystal, a surface instability emerges from the dipole in the unit cell,
independent of the surface instability naturally arising from dangling bond formation.
Even though Tasker's classification was introduced to describe the stability of ionic surfaces,
it is also relevant for covalent surfaces in the presence of finite ionicity.

Both dangling bond formation and the instability of polar surfaces are at play,
and they are reconciled simultaneously at the surface of covalent crystals.
Therefore, direct observation of the instability of polar surfaces in covalent systems has been difficult.
When we consider an epitaxial interface which has similar charge structure as the polar surface,
the instability from the dangling bonds disappears, and we can solely discuss the stability
in the same manner as that for the ideal ionic surfaces, as discussed in the next section.

\subsection{Polar semiconductor interfaces}

Similar to the instability of polar surfaces,
dipoles in the unit cells stacking from the interface
can cause potential divergence and instability, and require reconstruction.
This point was first proposed at the heteroepitaxial interface of GaAs and Ge in the [001] direction
by three groups from different starting points for treating ionicity.
Based on their considerations, we define polar and non-polar interfaces,
in analogy to polar and non-polar surfaces.

\subsubsection{Charge transfer based on electronegativity}

Frensley and Kroemer calculated the energy band diagram 
at abrupt semiconductor heterojunctions \cite{Frensley1977}.
Their starting point was to describe the alignment of atoms around the interface 
without considering the ionicity,
and then calculate the charge transfer based on the ionicity using the electronegativity of the atoms,
under assumption that the charge transfer only occurs between the nearest neighbors.
The Phillips electronegativity values $X_{\mathrm{Ph}}$ were used 
($X_{\mathrm{Ph}}(\mathrm{Ga}) = 1.13$, $X_{\mathrm{Ph}}(\mathrm{Ge}) = 1.35$, and
$X_{\mathrm{Ph}}(\mathrm{As}) = 1.57$ \cite{Phillips1973}).
In the bulk zincblende structure \textit{AB}, 
the \textit{A} site is tetrahedrally coordinated by four \textit{B} atoms and \textit{vice versa},
and based on their calculation \cite{Frensley1977APL}, the ionic charges $e^*$ of the atoms are given by 
\begin{equation}
e^* (A) = - e^* (B) = 0.76 q_0 \times [X_{\mathrm{Ph}}(B) - X_{\mathrm{Ph}}(A)], 
\end{equation}
where $q_0$ is the elementary charge.
This is equivalent to assuming a charge transfer of 
$0.76 q_0 \times \frac{1}{4} [X_{\mathrm{Ph}}(B) - X_{\mathrm{Ph}}(A)]$
between any pair of the nearest neighbors.

\begin{figure}[]
\begin{center}
\includegraphics[width = 14cm]{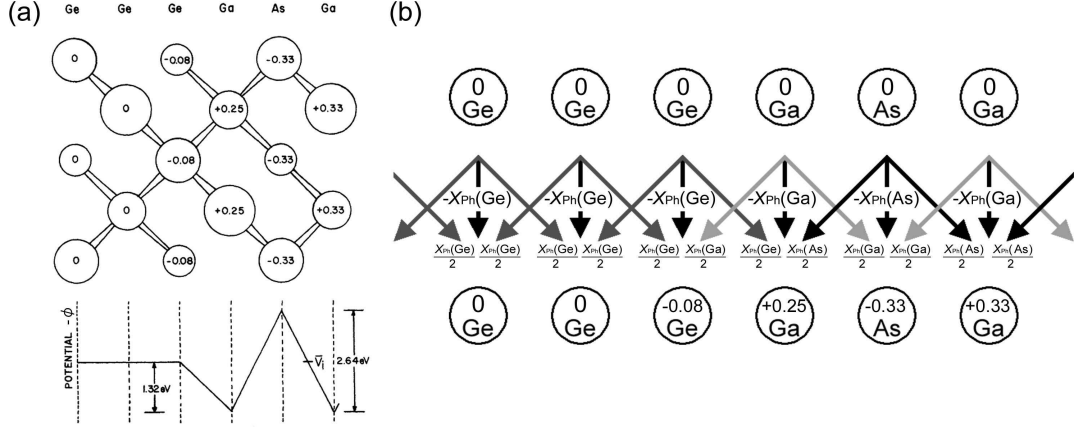}
\caption{\label{local-GaAsGeKroemer} 
(a) Model for a (001) Ge/GaAs heterojunction
considering the ionic character of the bonds.
The atomic positions and effective ionic charges are shown above.
Below is a diagram of the plane-averaged potential
(Frensley and Kroemer \cite{Frensley1977}).
(b) Schematic diagram of the charge transfer from the neutral atoms
with respect to the electronegativity.
}
\end{center}
\end{figure}

Consider the case of Ge/GaAs interfaces as shown in Fig.~\ref{local-GaAsGeKroemer}(a),
where the charge $e^* (\mathrm{Ga_{int}})$ on the Ga ions adjacent to the interface is
\begin{equation}
e^* (\mathrm{Ga_{int}}) =  0.76 q_0 \times \left[\frac{1}{2}X_{\mathrm{Ph}}(\mathrm{Ge}) + \frac{1}{2}X_{\mathrm{Ph}}(\mathrm{As}) - X_{\mathrm{Ph}}(\mathrm{Ga})\right] = 0.25q_0. 
\end{equation}
Similarly, the charges $e^* (\mathrm{Ge_{int}})$ on the Ge ions at the interface and 
$e^* (\mathrm{As})$ at the As sites are $e^* (\mathrm{Ge_{int}}) = -0.08q_0$ and 
$e^* (\mathrm{As}) = -0.33q_0$.

Without ionicity, the electrostatic potential is constant, 
and even with ionicity, the potential does not diverge.
This can be understood easily by tracking the virtual charge transfer processes 
from the starting alignment of the charge neutral atoms.
The charge transfer is equivalent to the situation that 
each neutral atom loses $ 0.76 q_0 \times X_{\mathrm{Ph}}$ charges,
and half of them are transferred to the left atoms and the other half to the other side, 
as shown in Fig.~\ref{local-GaAsGeKroemer}(b).
Therefore, charges are always transferred symmetrically, and each modulation creates no dipole,
resulting in no potential shift.
Here, a change of the number of electrons in the ions
compared to the bulk state is assumed,
which is equivalent to changing the valence assignments.

\subsubsection{Self-consistent calculation and counting electrons of bonds}

\begin{figure}[]
\begin{center}
\includegraphics[width = 14cm]{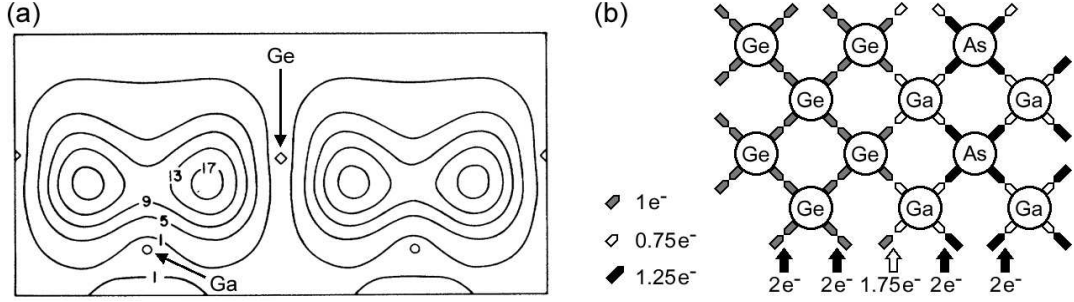}
\caption{\label{local-Barraf-GeGaAs} 
(a) Calculated contour plot of charge density for the partially occupied interface band around Ga-Ge bonds (Baraff \textit{et al.} \cite{Baraff}).
(b) Schematic model for counting electrons from the local-bond point of view.
}
\end{center}
\end{figure}

Baraff \textit{et al.} performed a self-consistent calculation 
of the potential, charge density, and interface states
for the abrupt interface between Ge and GaAs, terminated on a (001) Ga plane \cite{Baraff}. 
As shown in Fig.~\ref{local-Barraf-GeGaAs}(a),
their calculation showed fractional occupancy of electronic states 
only at the interface, which cannot exist in bulk.

These interface states can be discussed from a local-bond counting point of view as well.
The number of the valence electrons is 4 for Ge, 3 for Ga, and 5 for As.
When we assume all the valence electrons of an atom are equally distributed to the four covalent bonds
around it,
Ge, Ga, and As atoms supply 1, 0.75, and 1.25 electrons to each bond, respectively.
As shown in Fig.~\ref{local-Barraf-GeGaAs}(b),
the Ge-Ga bonds at the interface have 1.75 electrons per bond,
while other bonds have 2 electrons in each.
These partially occupied bonds are considered to form the interface states.

When the number of electrons are smaller than that in the bulk,
the attractive interaction between the bonded atoms should be weaker.
According to their calculation, 
the energy is minimized when the Ge-Ga bond length is 4 \% larger than that of the bulk.
Without the change of the bonding length, the system requires long-range disturbance of the lattice,
which is unlikely to be realized.

\subsubsection{Solving the potential divergence from the ionic picture}

Both the charge transfer model by Frensley and Kroemer and 
the electron counting model by Baraff \textit{et al.}
predict (indeed require) that there are interface states at a Ge/GaAs (001) interface even if it is perfectly abrupt, with no crystalline defects.
The central point raised by these studies is that despite having the same crystal structure, and having very closely matching lattice constants, this interface must accomodate charge arising from interface boundary conditions.
However, experimentally no considerable density of interface states was observed \cite{Grant1978},
and a model to treat this interface without changing the number of charges at the interface
was required.

\begin{figure}[]
\begin{center}
\includegraphics[width = 14cm]{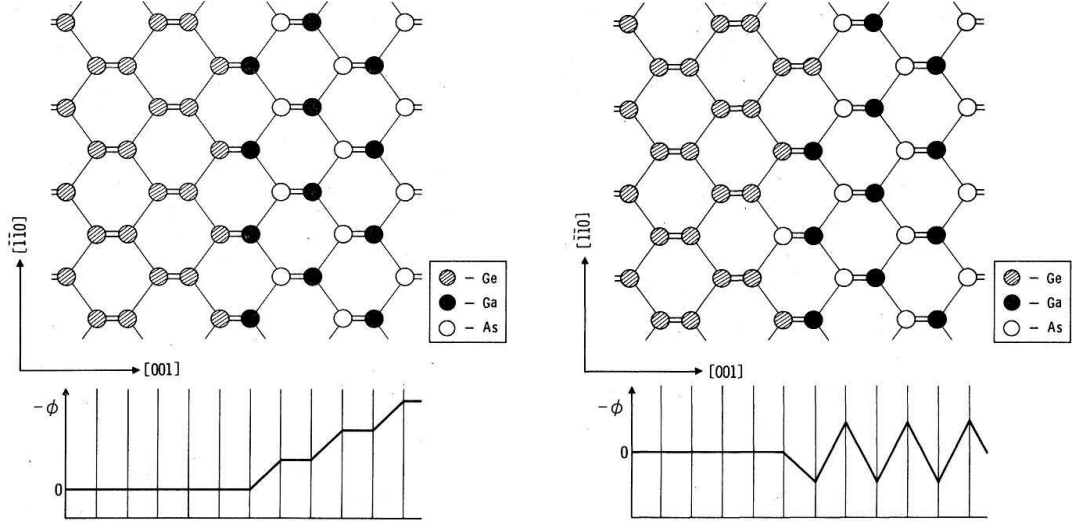}
\caption{\label{local-HarrisonGeGaAs} 
Schematic crystal structure
and electrostatic potential $\phi$
in the heterojunctions of Ge and GaAs in the [001] direction.
(a) An atomically abrupt interface.
(b) A Ge/GaAs heterojunction with two off-stoichiometric transition layers (Harrison \textit{et al.} \cite{Harrison}).
}
\end{center}
\end{figure}

Harrison \textit{et al.} constructed a model for the Ge/GaAs (001) heterojunctions
by stacking the fully ionized atoms,
and calculated the electrostatic potential based on the fixed assignment of the charges \cite{Harrison}.
As shown in Fig.~\ref{local-HarrisonGeGaAs}(a), the potential is very similar to the case of polar surfaces since the unit cells which \textit{start from the interface}
have dipoles in GaAs, while the unit cells in Ge are always charge neutral.
Therefore, the stacking of the dipoles causes potential divergence in this case as well.

Since the charge of each ion was fixed, 
the solution to the instability of this interface requires compensation
by changing the stoichiometry at the interface.
They proposed a simple model where 1/4 of the Ge atoms are replaced by As atoms at the interface
while 1/4 of the Ga atoms adjacent to the interface are replaced by Ge.
In this reconstructed model with two transition layers, the electrostatic potential
does not diverge, as shown in Fig.~\ref{local-HarrisonGeGaAs}(b).

It might be surprising that from two completely different starting points,
namely one from covalent (Frensley and Kroemer) 
and the other from ionic (Harrison \textit{et al.}) pictures,
exactly the same potential diagrams were obtained.
However, rearranging the number of charges at the perfectly abrupt interface
or changing the interface composition while maintaining the ionic charges of the atoms
can give the same net charge distribution.
The experimental absence of localized states suggests the atomic reconstruction based on the ionic picture.
This is equivalent of saying that the electronic state at this semiconductor heterointerface cannot deviate so strongly from that of the bulk constituents --- it is energetically inaccessible.  This is the fundamental aspect which can be quite different in complex oxide heterointerfaces, and is the subject of much current excitement. 
Namely, there is a possibility that the charge transfer picture (Frensley and Kroemer) and
the electron counting picture (Baraff \textit{et al.}), 
which require large deviations of electron numbers from the bulk values,
can be energetically accessed in transition metal oxides with multi-valency,
as described in Section \ref{sec-Metallic conductivity between two insulators}.

\subsubsection{Definition of the polar interface}

\begin{figure}[]
\begin{center}
\includegraphics[width = 15cm]{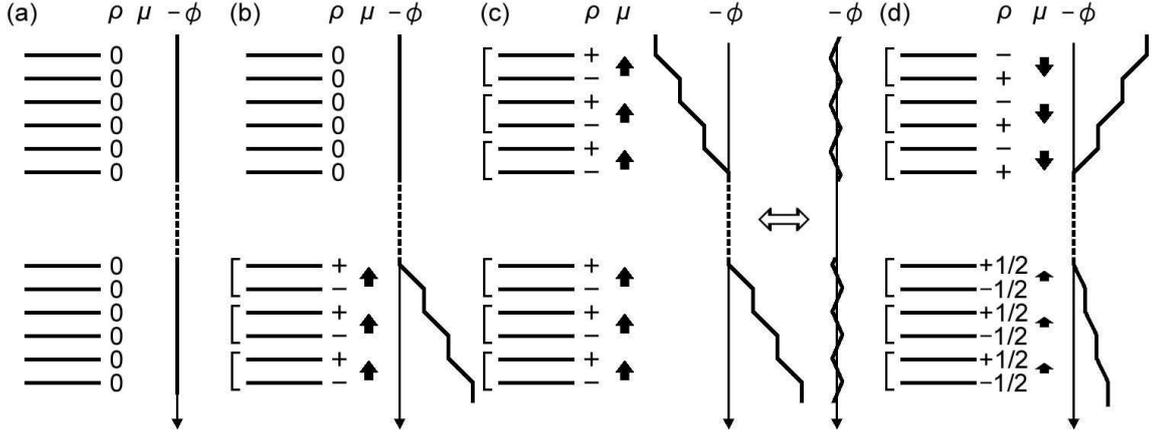}
\caption{\label{local-interface-def} 
Schematic distribution of charges $\rho$ on planes, the dipole moment $\mu$ in unit cells starting from the interface, and the electrostatic potential
for the four stacking sequences parallel to the interface,
(a) type I, (b) type II, (c) type III, and (d) type IV.
In order to treat a \textit{bulk frozen} interface
as a set of two \textit{bulk frozen} surfaces,
vacuum is inserted between them (dashed lines).
The electrostatic potential $\phi$ was calculated taking the vacuum as the potential reference
except for the right plot in (c), where a constant electric field was added 
to show the absence of macroscopic band bending at the interface.
}
\end{center}
\end{figure}

Following the model by Harrison \textit{et al.}, 
we can define the polar nature of a \textit{bulk frozen} interface between two materials,
where the interface consists of two \textit{bulk frozen} surfaces connected together\footnote{
Here, we consider only interfaces between two semi-infinite materials
--- we ignore the coupling to other interfaces or surfaces, which is discussed in Sections 
\ref{Coupling of polar discontinuities} and \ref{Modulation doping by a proximate polar discontinuity}.}.
First let us classify the interfaces by the polar nature of the two constituent surfaces, as shown in Fig.~\ref{local-interface-def}.

\begin{itemize}
\item Type I is formed between two non-polar surfaces, and the potential is flat.
\item Type II is formed between polar and non-polar surfaces, and the potential
diverges from the interface in the material with the polar surface.
\item Type III is formed between two polar surfaces, and the direction and the size of the dipoles
in the unit cells which start from the interface is the same.
Although $\phi$ looks to diverge from the interface 
in both materials,
there is no macroscopic difference in the potential slope across the interface.
We can cancel the potential slope on both sides
by adding a constant electric field, as shown in Fig.~\ref{local-interface-def}(c).
Therefore, this interface is stable as constructed.
\item Type IV is formed between two polar surfaces, 
where the dipoles in the two different unit cells are not identical, 
which results in a macroscopic difference in the potential slope at the interface.
It is impossible to find any constant electric field to cancel out the potential divergence 
in both media, due to this difference\footnote{
Note type IV is the most common and general case in reality,
since the electronegativity can never perfectly match between different materials.}.
\end{itemize}

Type I and type III are stable due to the absence of a macroscopic difference of the potential slopes,
while type II and type IV are not stable.
This difference arises from the continuity/discontinuity of the dipoles in the unit cells 
which start from the interface.
A \textit{bulk frozen} interface is non-polar, 
when the dipoles in the unit cells starting from the interface are identical
across the interface, and thus no change of the macroscopic potential slope exists.
On the other hand, it is polar
if a \textit{bulk frozen} interface has a discontinuity of the dipole moment in each unit cell.
For example, from a purely ionic viewpoint (Harrison \textit{et al.}) the abrupt Ge/GaAs (001) interface is classified as type II, and thus polar.

This definition of the polar nature of interfaces is consistent with that of surfaces.
When the vacuum is treated as a charge neutral medium,
the non-polar surface is type I, and the polar surface is type II in this classification of the interfaces,
and the polar nature is maintained following the definitions for the interface.
We can treat surfaces and interfaces in one framework,
which is the polar nature of discontinuities at materials boundaries.
In summary, polar discontinuities, which consist of polar surfaces and interfaces,
are unstable due to the macroscopic potential folding arising from the discontinuity of the stacking
of dipoles in the unit cells, and require reconstructions to stabilize them.
Interfaces with continuous polarity, on the other hand, are stable without any reconstructions.

\section{Metallic conductivity between two insulators}
\label{sec-Metallic conductivity between two insulators}

As noted in the introduction, the modern ability to approach atomic control in complex oxide heterostructures has newly enabled the experimental investigation of their polar discontinuities. 
As illustrated in Fig.~\ref{Perovskite}(a), an immediate question arises regarding the choice of the termination layer at the interface, and the consequences of this degree of freedom.
This issue has been most explicitly raised, and hotly debated, for the electron gas observed at the interface of two perovskite insulators, LaAlO$_3$ and SrTiO$_3$ \cite{Ohtomo04}.
Specifically, the (001) heterointerface was found to be insulating when grown using a SrTiO$_3$ substrate which was SrO-terminated, and conducting when TiO$_2$-terminated.
Given the rapid evolution of the field, and the numerous reviews of this heterostructure in the literature, we do not attempt a comprehensive review here. 
Instead the LaAlO$_3$/SrTiO$_3$ interface will be used to illustrate the various mechanisms suggested to explain the interface electronic structure, and the electrostatic boundary conditions which arise.  
It should be stressed, however, that all oxide heterointerfaces should be considered type IV to varying degrees, and thus these issues are quite general
-- even underlying the interface charge in the superlattice shown in Fig.~\ref{Perovskite}(c).

\subsection{The \textit{polar discontinuity} scenario}

\begin{figure}[]
\begin{center}
\includegraphics[width = 11cm]{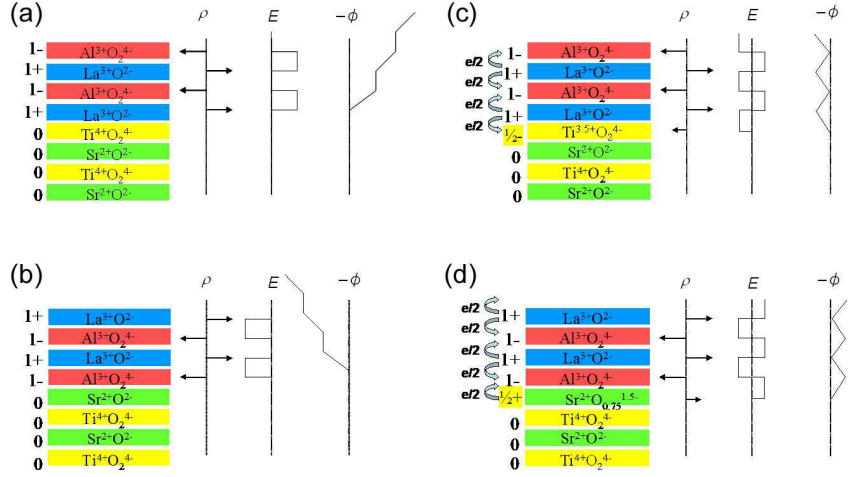}
\caption{\label{local-whysome} 
Polar reconstructions at the LaAlO$_3$/SrTiO$_3$ interfaces.
The unreconstructed (a) LaO/TiO$_2$ terminated interface,  
(b) AlO$_2$-SrO terminated interface,
(c) and (d) the corresponding reconstructed interfaces, respectively
(Nakagawa \textit{et al.} \cite{Nakagawa}).
}
\end{center}
\end{figure}

Assuming pure ionicity, the charge sequence of the (001) perovskite planes are different in these two materials,
namely the planes of LaAlO$_3$ are (La$^{3+}$O$^{2-}$)$^+$ and (Al$^{3+}$O$^{2-}_2$)$^-$,
while those of SrTiO$_3$ are (Sr$^{2+}$O$^{2-}$)$^0$ and (Ti$^{4+}$O$^{2-}_2$)$^0$.
Therefore, the abrupt interface between LaAlO$_3$ and SrTiO$_3$ is type II polar\footnote{
Allowing for covalency, SrO and TiO$_2$ planes in SrTiO$_3$ are no longer
charge neutral, and thus the (001) surface is weakly polar,
but still the dipole size of the unit cells 
which start from the interface is different from that of LaAlO$_3$.}
and requires reconstruction as shown in Fig.~\ref{local-whysome}, just as for the GaAs/Ge (001) interface.

Unlike polar semiconductor interfaces where only atomic reconstructions are available 
due to the fixed ionic charge of each element\footnote{While here we discuss large scale charge modifications, small polar discontinuities can induce free carriers in semiconductors, such as in AlGaN/GaN heterostructures \cite{Ambacher1999, Ibbetson2000}.},
we have another possibility to reconcile the instability of polar interfaces,
through electronic reconstructions \cite{Hesper2000}.
At the LaAlO$_3$/SrTiO$_3$ interface, 
when the interface termination is LaO/TiO$_2$,
it requires a net half negative charge per 2D unit cell to reconcile the potential divergence (n-type).
Accessing Ti$^{3+}$ can source this negative charge 
by accommodating electrons at the Ti 3$d$ level, as was spectroscopically observed \cite{Nakagawa}.
On the other hand, when the interface is terminated by AlO$_2$/SrO, 
the sign of the required charges is opposite (p-type). 
Due to the difficulty of accommodating holes in this structure (such as Ti$^{5+}$),
the positive charges are realized by the formation of oxygen vacancies close to the interface, as inferred from measurements of the O-K edge fine structure.
Oxygen vacancies are known as electron donors,
but in this case they are formed to provide positive charges,
and thus no electrons are supplied from these vacancies compensating the polar discontinuity.
Therefore the system does not have itinerant electrons and remains insulating \cite{Ohtomo04, Nishimura2004}.

One of the key corollaries of this scenario is the LaAlO$_3$ thickness dependence.
This is because the size of the potential difference arising from the stacking of the dipoles 
in the unreconstructed structure
is finite in thin films, and if it is small enough, the system may be stable without any reconstruction. 
Indeed, a critical thickness of LaAlO$_3$ was observed \cite{Thiel},
where the n-type LaAlO$_3$/SrTiO$_3$ interface is insulating
if the thickness of LaAlO$_3$ is up to 3 unit cells, and metallic above that thickness.
A similar tendency was also observed in SrTiO$_3$/LaAlO$_3$/SrTiO$_3$ trilayer structures 
where the distance between the two polar interfaces was varied \cite{Huijben}.

\subsection{Oxygen vacancy formation during growth}

SrTiO$_3$ is known to be a material which readily accommodates oxygen vacancies
that act as donors to provide itinerant electrons \cite{Tufte1967, Frederikse1967}.
Either kinetic bombardment of the SrTiO$_3$ substrate by the ablated species  (early studies of this interface all used pulsed laser deposition), or gettering by a reduced film, can induce oxygen vacancies, 
and they were suggested to be the dominant origin for the observed conductivity by several groups \cite{Kalabukhov, Siemons, Herranz}.
Indeed, the first report found a strong dependence of the Hall density for n-type interfaces on the oxygen partial pressure ($P_{{\rm O}_2}$) during growth, while the p-type interface was robustly insulating \cite{Ohtomo04}.
For n-type samples with similar variations in the transport properties,
Kalabukhov \textit{et al.} found when grown at $P_{{\rm O}_2}=10^{-6}$ Torr,
the samples exhibited blue cathode- and photo-luminescence at room temperature \cite{Kalabukhov},
similar to that of reduced SrTiO$_3$ by Ar bombardment \cite{Kan2005}.
In addition to transport studies, Siemons \textit{et al.} demonstrated that the photoemission spectra from these interfaces showed a larger amount of Ti$^{3+}$ in samples
grown at low pressures without oxygen annealing \cite{Siemons}.
Herranz \textit{et al.} observed Shubnikov-de Haas oscillations in reduced LaAlO$_3$/SrTiO$_3$ samples which were quite similar to bulk doped SrTiO$_3$, and rotation studies indicated a three-dimensional Fermi surface \cite{Herranz}.

\begin{figure}[]
\begin{center}
\includegraphics[width = 11cm]{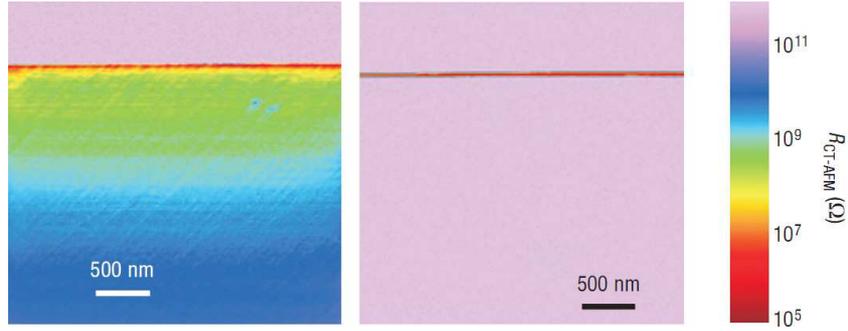}
\caption{\label{local-AFM} 
Conducting tip atomic force microscopy mapping around the LaAlO$_3$/SrTiO$_3$ interface of (a) the as grown sample ($P_{{\rm O}_2}=10^{-5}$ Torr) 
and (b) the postannealed sample (Basletic \textit{et al.} \cite{Basletic2008}).}
\end{center}
\end{figure}

The strong $P_{{\rm O}_2}$ dependence of the conducting channels in LaAlO$_3$/SrTiO$_3$ was observed and mapped by means of conducting tip atomic force microscopy on cross-sections of the interface, which revealed a conducting region extending $> 1 ~\mathrm{\mu m}$ into the substrate for samples grown at low pressure \cite{Basletic2008}, as shown in Fig.~\ref{local-AFM}.
After annealing, the width of the conductive layer decreased to $\sim 7 ~\mathrm{nm}$,
as limited by the radius of the probing tip.
The consensus of these and other studies was that the free carriers in samples grown at low $P_{{\rm O}_2}$ were dominated by oxygen vacancies, since the density far exceeded that needed to stabilize the polar discontinuity.  For high $P_{{\rm O}_2}$, or after post-annealing, the origin was less clear.

\subsection{Intermixing and local bonding at the interface}

\begin{figure}[]
\begin{center}
\includegraphics[width=12cm]{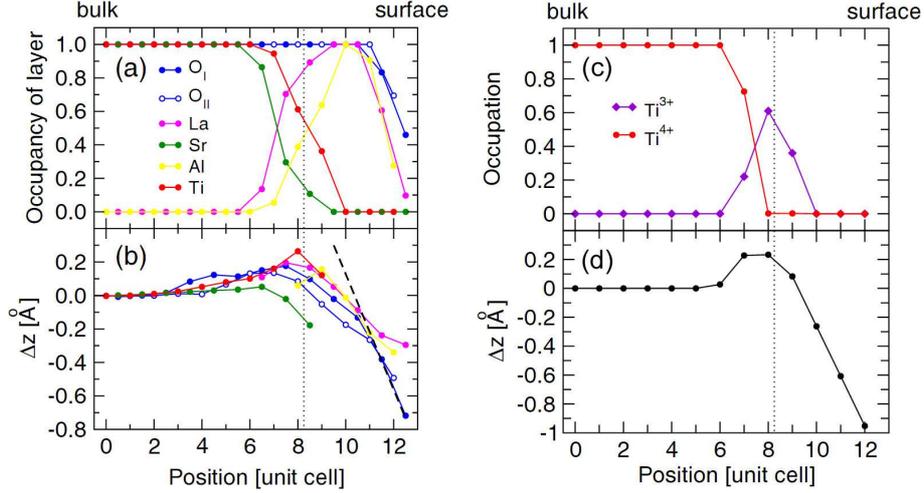}
\caption{\label{local-Willmott} 
(a) Occupancies and (b) cumulative displacements $\Delta _z$ of the atoms at the LaAlO$_3$/SrTiO$_3$
interface. (c) Concentration of Ti$^{3+}$ determined by a minimization of the electrostatic potential. (d) Predicted cumulative unit cell displacements from bulk positions, based on the component ionic radii (Willmott \textit{et al.} \cite{Willmott2007}).}
\end{center}
\end{figure}

Willmott \textit{et al.} studied a five unit cell film of LaAlO$_3$ on SrTiO$_3$ (001)
by means of surface x-ray scattering \cite{Willmott2007},
using coherent Bragg rod analysis (COBRA) \cite{Yacoby2003} and further structural refinement.
Their analysis revealed intermixing of the cations (La, Sr, Al, and Ti) at the interface [Fig.~\ref{local-Willmott}(a)],
as well as significant local displacement of the atomic position both in the film, and in SrTiO$_3$
close to the interface [Fig.~\ref{local-Willmott}(b)].
The distribution of Ti valence was also inferred by minimizing the electrostatic potential [Fig.~\ref{local-Willmott}(c)]
following the obtained atomic positions,
and the atomic displacements were explained by the larger ionic radii of Ti$^{3+}$ compared to Ti$^{4+}$ [Fig.~\ref{local-Willmott}(d)]. 
Based on these observations, the origin of the interface conductivity was suggested to be the formation of the bulk-like solid solution La$_{1-x}$Sr$_x$TiO$_3$ in a region of approximately 3 unit cells.  
This explanation can be considered a diffused version of local bonding arguments --- i.e., that even in the abrupt limit, the Ti at the interface has La on one side, and Sr on the other.

\subsection{Reconciling the various mechanisms}

At present, it can be fairly stated (we believe) that no one scenario can completely explain the vast and growing body of experimental work on this system.  
Even theoretical calculations show an extreme sensitivity to the choice of boundary conditions and assumptions of site-occupancy \cite{Park2006, Popovic2008, Lee2008, Ishibashi2008, Bristowe2009}.
To give examples for each perspective: The polar discontinuity picture should lead to a significant internal field in ultrathin LaAlO$_3$, 
while experiments \cite{Segal2009} put an upper bound far below that expected theoretically, even allowing for significant lattice polarization \cite{Pentcheva2009}.  
Oxygen vacancies induced by growth are difficult to reconcile with the notion that a single layer of SrO can prevent their formation.  
Local bonding and interdiffusion considerations do not address the constraints of global charge neutrality. 
Furthermore, discriminating between these mechanisms is often difficult, since the change of Ti valence shows similar transport, spectroscopic, and optical properties, independent of its origin.

It is extremely likely that multiple contributions exist to varying degrees, dependent on the growth details of a given sample.  
While this is a matter for further experimental investigation and refinement, an equally difficult issue appears to be one of semantics. For example, 
one of the conceptual difficulties of the \textit{polar discontinuity} picture 
has been the question: 
Where do the electrons come from?
In many presentations \cite{Nakagawa, Thiel, Levy, Takizawa2009}, 
the charges at the interface are depicted to originate from the surface of the LaAlO$_3$ film,
but it is not so obvious that they travel through the insulating films independent of its thickness \cite{Thiel}.
Fundamentally, the ``non-locality'' of these electrostatic descriptions has sometimes been considered less intuitive and compelling than ``local chemistry'' mechanisms such as vacancies or interdiffusion \cite{Kalabukhov, Siemons, Basletic2008, Yoshimatsu, Willmott2007}.

A related difficulty is how to treat the charge density to describe the macroscopic electric field.
Based on the \textit{polar discontinuity} picture,
the stacking of the dipoles in the unit cells creates a macroscopic electric field,
resulting in the change of the potential slope at the discontinuity.
In this picture, the unit cells start from the discontinuity, 
and thus the composition is always the same as that of the bulk, which is charge neutral.
It might be strange that the potential starts to bend at the discontinuity
although all the unit cells start out charge neutral,
because the source of an electric field is charge.
In fact, the source is the macroscopic bound charge density at the interface, as discussed in Section \ref{sec-Equivalence of the two pictures},
but the existence of such implicit charge density has caused a fair bit of confusion.
To address these concerns, it is useful to treat
the electrostatics in a purely local description, as well as the boundary charges explicitly.
Furthermore, the effects of defects and diffusion can be discussed more simply by re-framing the polar discontinuity picture in local form.  Therefore, this \textit{local charge neutrality} picture based on dipole-free unit cells 
is developed in the next section
first for idealized models, followed by discussion of incorporation of chemical defects.

\section{The \textit{local charge neutrality} picture}

\subsection{Unit cells in ionic crystals}\label{sec-Unit cells in ionic crystals}

\begin{figure}[]
\begin{center}
\includegraphics{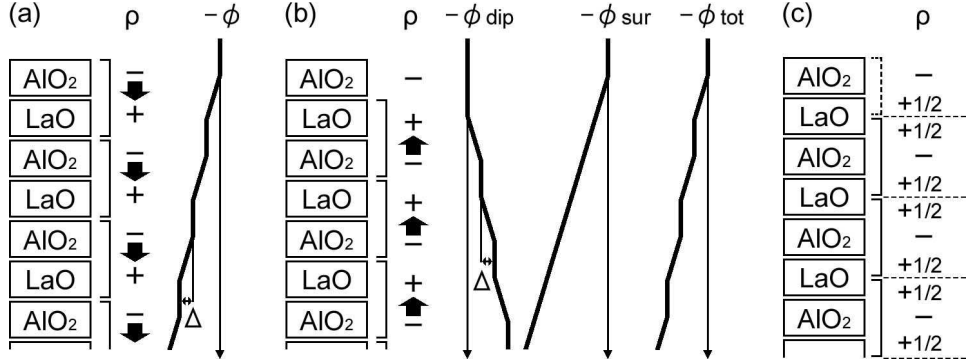}
\caption{\label{local-1} 
Schematic illustrations of the various choices of unit cells in LaAlO$_3$, charge distribution $\rho$, and associated potential $\phi$.
The filled allows indicate the orientation of the dipoles with size $\Delta$ in the unit cells.
(a) Taking [(AlO$_2$)$^-$ (LaO)$^+$] as a unit cell.
(b) Taking [(LaO)$^+$ (AlO$_2$)$^-$] as a unit cell, and the total potential $\phi_{\mathrm{tot}}$
is the sum of the potential $\phi_{\mathrm{dip}}$ arising from the stacking of the dipoles
and the potential $\phi_{\mathrm{sur}}$ from the surface charge.
(c) Taking a dipole-free unit cell [$\frac{1}{2}$(LaO) - (AlO$_2$) - $\frac{1}{2}$(LaO)].
}
\end{center}
\end{figure}

One of the origins of confusion regarding the stability and reconstructions 
of polar discontinuities arises
because of the various choices of a unit cell in a crystal,
which determines the size and direction of the dipole in it\footnote{For simplicity, we started our discussion from a point-charge model of ions, and neglect the free carrier distribution. More generally, the charge distribution can be described using contributions from ion cores and free carrier Wannier functions, and the ambiguity of the choice of unit cells appears in this extended case as well \cite{Massimilliano2009}. }.
For example, when we take the unit cell of LaAlO$_3$, this stacking can be treated as a dipole of [(AlO$_2$)$^-$ (LaO)$^+$], as shown in Fig. \ref{local-1}(a).
However, it is also possible to take [(LaO)$^+$ (AlO$_2$)$^-$] as a unit cell, 
and the sign of the dipole is opposite to the previous case, as shown in Fig.\ref{local-1}(b).

The choice of unit cells should not change the electrostatic potential in the crystal.
Indeed, the difference between the two choices is compensated 
by the potential $\phi_{\mathrm{sur}}$ arising from the surface layer.
If the surface is terminated by the (AlO$_2$)$^-$ layer, it remains as an extra negatively charged layer 
when we take [(LaO)$^+$ (AlO$_2$)$^-$] as the unit cell, as shown in Fig.\ref{local-1}(b),
and the total electrostatic potential remains the same as in the [(AlO$_2$)$^-$ (LaO)$^+$] unit cell case.

Therefore, it is impossible to fix the direction of the dipoles
without knowing the surface termination.
In other words, it is the surface that determines the stability of the system.
So the problem can be simplified by considering the surface locally,
and not by counting the number of dipoles in the material.
In order to avoid the dipoles arising from the stacking of charged layers,
the simplest approach is to take dipole-free unit cells.
This can be achieved in any crystal \cite{Goniakowski2008},
and in the LaAlO$_3$ case,
this is done by taking  [$\frac{1}{2}$(LaO) - (AlO$_2$) - $\frac{1}{2}$(LaO)] 
(or [$\frac{1}{2}$(AlO$_2$) - (LaO) - $\frac{1}{2}$(AlO$_2$)])
as a unit cell, as shown in Fig. \ref{local-1}(c).
This is analogous to the unit cell in a type 2 model in Tasker's classification.
From group theory, it is known that spontaneous polarization
can be observed only in a direction where the crystal does not have mirror symmetry.
Since cubic perovskites do have mirror symmetry in the [001] direction (we neglect surface or interface induced lattice polarization for now), 
it is useful to take dipole-free unit cells to reflect the lack of polarization in the bulk.

When we take this unit cell, the stability of a polar surface or interface
can be discussed by only considering the charge neutrality of each unit cell,
since now there is no net dipole created by the stacking of charged layers.
For example, the instability of the (AlO$_2$) terminated surface of LaAlO$_3$ is naturally derived
because the top-most unit cell is [(AlO$_2$)$^-$ $\frac{1}{2}$(LaO)$^+$]$^{-0.5}$, 
which violates charge neutrality, as shown in Fig. \ref{local-1}(c).
Thus the (001) surface of LaAlO$_3$ cannot keep the bulk termination, either by an AlO$_2$ or LaO layer,
and must reconstruct to compensate this charge \cite{Francis, Lanier}.
Once charge neutrality of all the unit cells is achieved,
the system is free to undergo
electron / hole modulation or interdiffusion / displacement of the atoms,
which creates only finite dipoles and does not violate neutrality\footnote{
The effect of diffusion is further discussed in Section \ref{sec-Effect of diffusion} as an example of sources of such finite dipoles.
Another source of interface dipoles, the \textit{quadrupolar discontinuity}, is discussed in Section \ref{sec-quadrupolar discontinuity}.}.
These perturbations are important 
because they determine the real charge structure,
for example via lattice distortion close to the surface of LaAlO$_3$ \cite{Francis, Lanier}.

Following these arguments, the polar nature of 
given \textit{bulk frozen} surfaces and interfaces is clearly defined,
considering local charge neutrality using dipole-free unit cells:
a \textit{bulk frozen} surface or interface is polar 
if the unit cell at the surface or interface cannot keep charge neutrality
when we take dipole-free unit cells in the bulk.

\subsection{LaAlO$_3$/SrTiO$_3$ 
in the \textit{local charge neutrality} picture}

\begin{figure}[]
\begin{center}
\includegraphics[width=14cm]{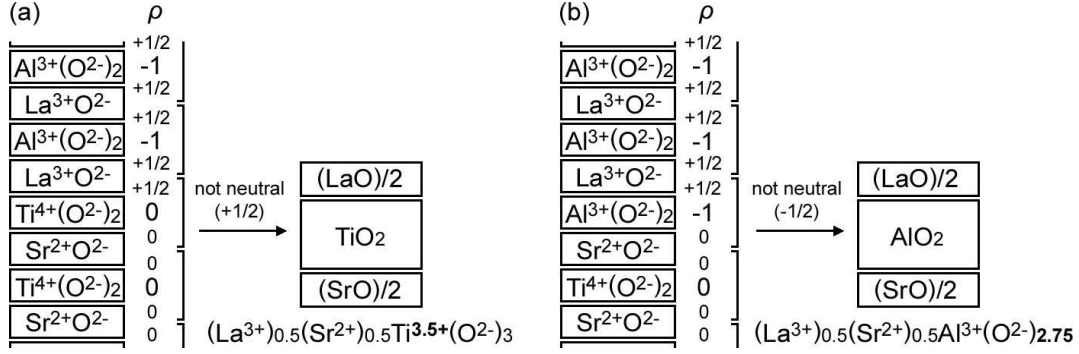}
\caption{\label{local-chargeLAOSTO} 
Schematic illustrations of the charge structure across the two types of \textit{bulk-frozen} 
LaAlO$_3$/SrTiO$_3$ interfaces,
assuming the ionic charges based on the valence states in bulk LaAlO$_3$ and SrTiO$_3$.
Taking dipole-free unit cells, the interface unit cell cannot keep charge neutrality,
and the simplest neutral interface stoichiometry is written in the right.
(a) LaO/TiO$_2$ terminated LaAlO$_3$/SrTiO$_3$ interface and
(b) AlO$_2$/SrO terminated LaAlO$_3$/SrTiO$_3$ interface.
}
\end{center}
\end{figure}

Taking dipole-free unit cells of perovskites $AB$O$_3$ in the [001] direction, 
namely [$\frac{1}{2}$($A$O) - ($B$O$_2$) - $\frac{1}{2}$($A$O)],
the interface unit cell can be treated as a $\delta$-dopant at the interface.
As shown in Fig.~\ref{local-chargeLAOSTO},
the interface unit cell does not keep the stoichiometry of either of the bulk materials,
not even a simple mixture of them.
This issue is actually one of the central opportunities of the interface science of heterostructures.
Namely, the LaO/TiO$_2$ terminated LaAlO$_3$/SrTiO$_3$ interface
has La$_{0.5}$Sr$_{0.5}$TiO$_3$ as the interface unit cell.
Considering the formal electronic charges of La$^{3+}$, Sr$^{2+}$, and O$^{2-}$,
the Ti ion in this unit cell should take Ti$^{3.5+}$ as a formal valence,
which is the same reconstructed state as that for the previous discussion
by the \textit{polar discontinuity} picture.

Similarly, the AlO$_2$/SrO terminated interface has
La$_{0.5}$Sr$_{0.5}$AlO$_3$ as the interface unit cell,
and due to the fixed ionic charges of the elements,
such a unit cell is not charge neutral and thus unstable: (La$^{3+}_{0.5}$Sr$^{2+}_{0.5}$Al$^{3+}$O$^{2-}_3$)$^{0.5-}$.
Instead, allowing a change of the oxygen number in the interface unit cell,
La$_{0.5}$Sr$_{0.5}$AlO$_{2.75}$ is charge neutral\footnote{This composition is not stable in bulk perovskite form, but can be considered as a mixture of the bulk compounds, LaAlO$_3$, Sr$_3$Al$_2$O$_6$, and SrAl$_2$O$_4$, stabilized epitaxially at the interface.},
and the decrease of the oxygen content at the interface
to stabilize a p-type LaAlO$_3$/SrTiO$_3$ interface is naturally derived.
Note in this case, these oxygen vacancies are introduced to keep charge neutrality,
and thus do not provide any free electrons.
Thus, the electrons/oxygen vacancies to reconcile the polar instability of
the \textit{bulk frozen} LaAlO$_3$/SrTiO$_3$ interfaces
are supplied by the interface unit cell itself.

\begin{figure}[]
\begin{center}
\includegraphics[width=10cm]{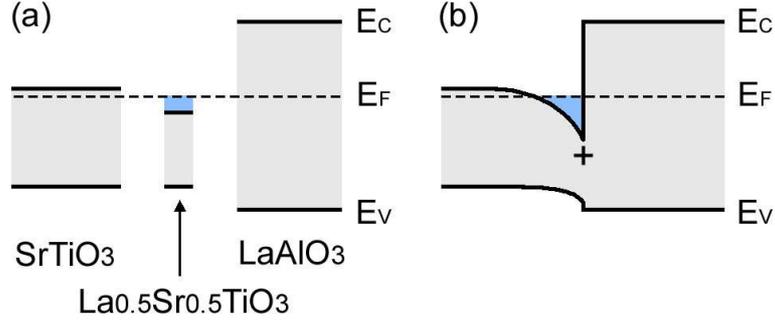}
\caption{\label{local-delta-LAOSTO} 
(a) Schematic band structures of SrTiO$_3$, LaAlO$_3$ and La$_{0.5}$Sr$_{0.5}$TiO$_3$,
assuming that SrTiO$_3$ and LaAlO$_3$ are intrinsic semiconductors.
(b) Schematic energy band diagram of the LaAlO$_3$/SrTiO$_3$ interface
with the $\delta$-dopant La$_{0.5}$Sr$_{0.5}$TiO$_3$ at the interface.}
\end{center}
\end{figure}

In semiconductors, $\delta$-doping is usually achieved in a symmetric geometry \cite{Gossmann1993},
that is the dopant layer is sandwiched between the same host material.
In the LaAlO$_3$/SrTiO$_3$ case, by contrast, the two sandwiching materials have different
band structure, with SrTiO$_3$ having the narrower bandgap.
Therefore, broadening of the charge distribution from the $\delta$-dopant
occurs only in the SrTiO$_3$ side \cite{Yoshimatsu}.
Figure~\ref{local-delta-LAOSTO} shows how the LaAlO$_3$/SrTiO$_3$ interface
can be depicted in a semiconductor energy band diagram.

\subsection{Coupling of polar discontinuities}
\label{Coupling of polar discontinuities}

\begin{figure}[]
\begin{center}
\includegraphics[width=11cm]{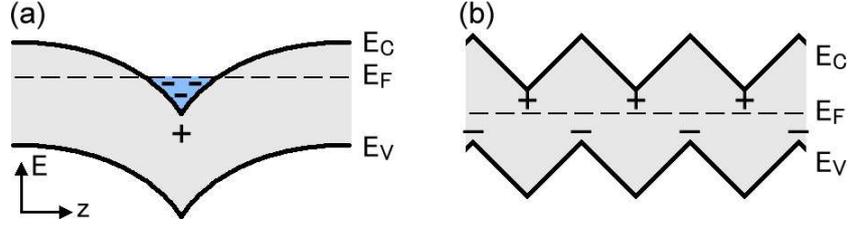}
\caption{\label{local-delta-proximity} 
Schematic band diagram of an (a) undepleted and (b) depleted $\delta$-doped semiconductor.
The undepleted structure contains free carriers as well as ionized impurities.
The depleted structure contains the same amount of donor and acceptor impurities (Gossmann and Schubert \cite{Gossmann1993}).
}
\end{center}
\end{figure}

When two polar discontinuities are brought in proximity to one another,
coupling of the charges can occur to minimize the total energy of the system.
This is just like the coupling of dopant layers in $\delta$-doped semiconductor heterostructures, which can be understood in terms of depleted and undepleted structures.
Figure~\ref{local-delta-proximity}(a) shows the band diagram of an undepleted semiconductor
with one layer of $\delta$-dopant.
Since the dopant is positively ionized, an equal number of free electrons are left,
and they screen out the potential created by the $\delta$-dopant.
As a consequence, the structure is neutral, and has zero electric field
sufficiently far away from the $\delta$-dopant layer.

On the other hand, Fig.~\ref{local-delta-proximity}(b) shows 
the band diagram of a depleted $\delta$-doped structure,
where the number of donors is equal to that of acceptors,
and they are sufficiently close in space.
As a result, all the free carriers recombine and the structure is depleted.
The critical parameters to treat electronic coupling of two $\delta$-doping layers
are the distance between them and the dielectric constant of the medium.
When the distance between them is smaller 
than the length scale of band bending of the undepleted structure,
they couple and the system goes to a depleted state.
Note in this case, charge neutrality in the neighborhood of one dopant layer
is not necessarily maintained.
Superlattice calculations using density functional theory
show that the above threshold is observed by changing the thickness of each layer in LaAlO$_3$/SrTiO$_3$ superlattices,
which can be captured in terms of a simple capacitor model \cite{Ishibashi2008, Bristowe2009}.

\begin{figure}[]
\begin{center}
\includegraphics[width=5cm]{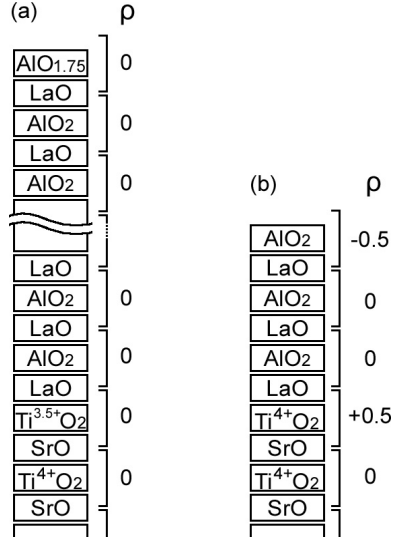}
\caption{\label{local-thickness-dep} 
Schematic illustration of the charges ($\rho$) of the dipole-free unit cells of 
atomically abrupt LaAlO$_3$/SrTiO$_3$ interfaces
in (a) the thick LaAlO$_3$ limit with the polar discontinuities locally neutralized, and 
(b) the thin LaAlO$_3$ limit with depleted polar discontinuities. 
}
\end{center}
\end{figure}

In $\delta$-doped semiconductor heterostructures, only considering modulation of free carriers
is sufficient to describe the coupling of the $\delta$-dopant layers.
However, when the system allows possibilities of excess charges other than the free carriers,
which come from outside of the constructed crystal --
e.g., anion or cation vacancies or foreign atoms absorbed to the surface,
the electrostatic potential can be minimized via them.
For example, the instability of the polar AlO$_2$-terminated LaAlO$_3$ surface can be solved by introducing 
positively charged surface oxygen vacancies \cite{Lanier}.
The LaAlO$_3$ thickness dependence of LaO-TiO$_2$ terminated LaAlO$_3$/SrTiO$_3$ can be explained
by a simple assumption, 
where we only consider coupling of the free electrons provided by the LaAlO$_3$/SrTiO$_3$ interface 
and the surface oxygen vacancies to keep the local charge neutrality of the polar LaAlO$_3$ surface.
Figure \ref{local-thickness-dep} shows a schematic structure of the LaAlO$_3$/SrTiO$_3$
heterojunction, where two polar discontinuities exist at the LaAlO$_3$/SrTiO$_3$ interface
and the LaAlO$_3$ surface.
When the LaAlO$_3$ film is sufficiently thick [Fig.~\ref{local-thickness-dep}(a)]
these polar discontinuities are decoupled and charge neutrality is preserved locally
by introducing oxygen vacancies at the surface and taking the Ti valence of $3.5+$ at the interface unit cell.
Fractionally filled Ti valence provides itinerant electrons, 
and therefore the system is metallic in the thick limit.

If the two polar discontinuities are brought closer [Fig.~\ref{local-thickness-dep}(b)], 
the external charges and the free carriers
(surface oxygen vacancies and extra electrons in Ti valence) 
can recombine in an environment with oxygen gas,
and the system is depleted and the conductivity disappears\footnote{
This is similar to the recombination of free electrons and holes,
and the actual recombination can be written by following Kr\"{o}ger-Vink notation \cite{Kosuge1993} as:
$\frac{1}{2} \mathrm{O}_2 + 2e' + \mathrm{V_O}^{\cdot \cdot} \rightarrow 0$.}.
Here, the word ``deplete'' is used to mean 
``reduce the amount of charge other than charge bound to the crystal'',
and the full depletion of the LaAlO$_3$ surface means extinguishing 
the positively charged oxygen deficiency at the surface 
--- i.e. the surface turns to an unreconstructed \textit{bulk frozen} state.
Manipulating this transition appears to roughly capture the essence of writing nanoscale features \cite{Levy}, which corresponds to the ``writing'' of surface charge \cite{Xie2010}.

\subsection{Modulation doping by a proximate polar discontinuity} 
\label{Modulation doping by a proximate polar discontinuity}

\begin{figure}[]
\begin{center}
\includegraphics[width=7.5cm]{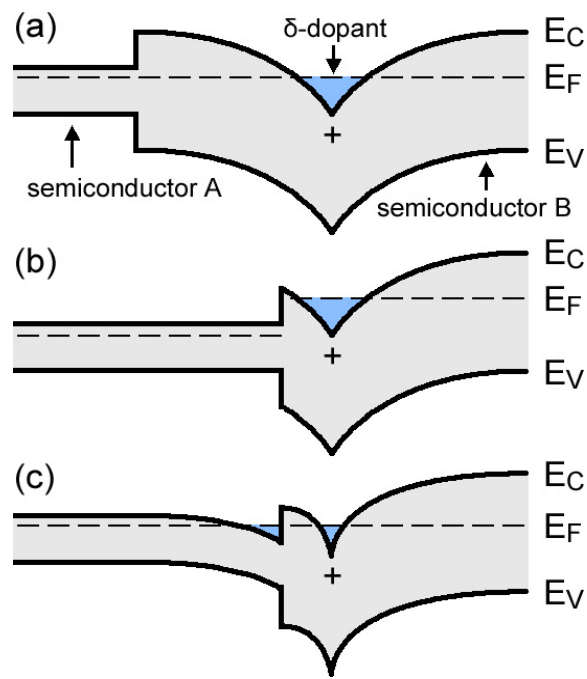}
\caption{\label{local-modulation} 
Schematic band diagram of the interface between two intrinsic semiconductors A and B,
with a layer of $\delta$-dopant in B.
(a) A is sufficiently far from the $\delta$-dopant in B. 
(b) Model with no charge modulation
though the distance between A and the $\delta$-dopant is close,
resulting in the mismatch of the chemical potential.
(c) Charges transfer from B to A, so as to match the chemical potential.
}
\end{center}
\end{figure}

Even in systems containing only one polar discontinuity,
coupling between the polar discontinuity and the other layers can occur
in analogy to modulation doping by a $\delta$-doping layer \cite{Dingle1978},
as shown in Fig.~\ref{local-modulation}.
At the semiconductor interface, lineup of the conduction and valence bands
should be maintained, as well as a fixed chemical potential,
which causes the modulation of carriers resulting in band bending.

Assume an interface between
two intrinsic semiconductors A (narrow bandgap) and B (wide bandgap).
If the $\delta$-dopant in B is sufficiently far from A,
the flat band condition at the interface is maintained [Fig.~\ref{local-modulation}(a)].
When A is brought in proximity to the $\delta$-dopant in B,
in order to keep the conduction band lineup,
the conduction band minimum of A lies 
at lower energy than the chemical potential in B [Fig.~\ref{local-modulation}(b)].
Since the free carriers around the $\delta$-dopant in B have higher energy
than the conduction band minimum of A,
they transfer to A and band bending occurs in A [Fig.~\ref{local-modulation}(c)].
As a result, the $\delta$-dopant in B is depleted, and A is doped close to the interface.

\begin{figure}[]
\begin{center}
\includegraphics[width=12cm]{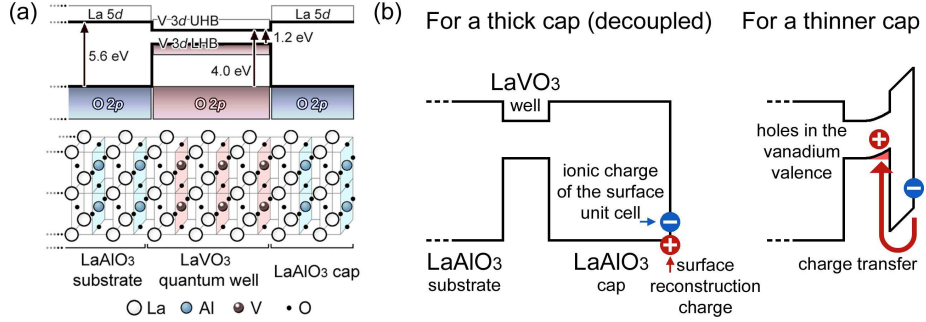}
\caption{\label{Modulation} 
(a) Schematic band diagram and crystal structure of a LaVO$_3$
quantum well embedded close to an AlO$_2$-terminated LaAlO$_3$ (001) surface. 
(b) Illustrations showing how reconstruction charge at the LaAlO$_3$ surface is transferred to the buried LaVO$_3$ quantum well. In order to solve the instability caused by the polar nature of AlO$_2$-terminated surface, positive charge is required. When the LaAlO$_3$ cap is sufficiently thick (left), 
the LaAlO$_3$ surface and the LaVO$_3$ well layer are separated, and the positive compensating charge remains at the surface. 
For a thinner spacing (right), the LaVO$_3$ well layer accommodates this positive charge, which is energetically more favored \cite{Higuchi2009, Takizawa2009}.}
\end{center}
\end{figure}

An experimental example of such charge modulation was observed 
from a type 3 polar surface of LaAlO$_3$ (001), which acts as a $\delta$-dopant,
to a narrow layer of the Mott insulator LaVO$_3$ with smaller bandgap, 
as shown in Fig.~\ref{Modulation} \cite{Higuchi2009, Takizawa2009}.
This system is noteworthy in the discussion of polar discontinuity effects, in that the observed hole-doping can neither arise from oxygen vacancies nor interdiffusion.

\subsection{Advantages of the \textit{local charge neutrality} picture}
\label{Advantages of the local charge neutrality picture}

As discussed, the \textit{local charge neutrality} picture gives a clear and self-consistent explanation
for the various phenomena at surfaces and interfaces.
The stability of given surfaces and interfaces can be simply judged
by looking at the local composition.
When the interface composition differs from that of the bulk or a simple superposition of them,
a different electronic and/or atomic state can be expected.
Enforcing local charge neutrality at the interface unit cell
naturally explains the source of the charges which can change the stoichiometry 
or the electron number of the constituent atoms from those of the bulk.
For example, in a LaO/TiO$_2$ terminated LaAlO$_3$/SrTiO$_3$ \textit{bulk frozen} interface,
the interface unit cell is La$_{0.5}$Sr$_{0.5}$TiO$_3$ and the Ti is $3.5+$
to achieve charge neutrality.
Therefore, the difference between the Ti$^{4+}$ in the bulk SrTiO$_3$ 
and the interface Ti$^{3.5+}$ comes from the interface unit cell itself,
and not from anywhere else.

Another advantage of taking the \textit{local charge neutrality} picture 
becomes clear when we discuss the coupling of polar discontinuities.
The stability of the system can be discussed through the distance between the polar discontinuities 
and the screening length of the host material.
In other words, the stability is determined by the balance of
the activation energy of the dopant
and the electrostatic energy allowing polarization of the media between the dopant layers.
Therefore, the total polarizability of the media can be considered in the calculation,
and is connected to the bulk permittivity in the thick limit.
Note when the media is thin, the local effective permittivity is non-trivial
since the local atomic displacements can be different from that in the bulk, and the local dielectric approximation breaks down \cite{Stengel2006}.

\section{Equivalence of the two pictures}\label{sec-Equivalence of the two pictures}

Thus far we have discussed boundary conditions based on two different choices for the unit cell. This was implicitly taking a microscopic viewpoint, since the electric field ${\bf E}$ and the total charge density $\rho_{\rm tot}$ were connected by Gauss' law: $\varepsilon _0 \nabla \cdot \mathbf{E} = \rho_{\mathrm{tot}}$,
where $\varepsilon_0$ is the vacuum permittitivity. Since $\rho_{\rm tot}$ is used (hence the atomic-scale stepped or sawtooth potentials), different choices for the unit cell were irrelevant so long as global charge neutrality was considered.  
Here we confirm the equivalence of the two pictures from the macroscopic electrostatic viewpoint.

\subsection{Gauss' law for infinite crystals}

In media, treating $\rho_{\rm tot}$ is often quite complicated, which can be simplified by using the electric displacement ${\bf D}$:
\begin{equation}
\mathbf{D} = \varepsilon_0 \mathbf{E} + \mathbf{P} \label{DEP},
\end{equation}
where ${\bf P}$ is the polarization. Then Gauss' law is given by 
\begin{equation}
\nabla \cdot \mathbf{D} = \rho_{\mathrm{free}}, \label{divD}
\end{equation}
where $\rho_{\mathrm{free}}$ is the free charge
-- the part of the macroscopic charge density due to excess charge not intrinsic to the medium, which are designated as bound charge $\rho_{\mathrm{bound}}$.
These definitions do not depend on whether the charges are localized or itinerant,
and are just introduced for practical convenience
to treat the displacement of the bound charges 
as the dielectric response of the media to the electric field.
Since the total charge is conserved ($\rho_{\rm tot}=\rho_{\rm free}+\rho_{\rm bound})$, by taking the divergence of Eq.~\eqref{DEP}, 
\begin{equation}
\rho_{\rm bound}= - \nabla \cdot {\bf P}. \label{divP}
\end{equation}
In infinite crystals, only the divergence of $\mathbf{E}$, $\mathbf{P}$, and $\mathbf{D}$
has physical meaning,
and the polarization arising from the density of unit cell dipole moments ${\bf P}_{\rm dipole}$ can be neglected
since it merely adds a constant value to $\mathbf{P}$.

\subsection{Gauss' law for finite crystals}

When the crystal is finite, ${\bf P}_{\rm dipole}$ drops to zero at the surface.
Thus the choice of unit cell is important,
since it changes the nature of the discontinuity at the surface.
According to Eq.~\eqref{divP}
the magnitude of the polarization discontinuity at the surface
is determined by $\rho_{\rm bound}$ there.
Since $\rho_{\rm tot}$ at the surface does not depend on the choice of unit cell,
uncertainty in the dipole moment of the unit cell
is compensated by whether the charges at the surface are defined to be free or bound.
--- the surface charge is bound when it belongs to a bulk unit cell,
and free when not, and this definition does not depend on the origin of the charges \cite{Mermin}.

For simplicity, let us adopt the bulk dielectric constant $\varepsilon$
to connect $\mathbf{E}$ 
and the induced polarization $\mathbf{P}_{\mathrm{ind}}$ by the field, namely 
\begin{equation}
\mathbf{P}_{\mathrm{ind}} = (\varepsilon - \varepsilon_0) \mathbf{E}, \label{Pind} 
\end{equation}
hence
\begin{equation}
\mathbf{P} = \mathbf{P}_{\mathrm{ind}} + \mathbf{P}_{\mathrm{dipole}}. \label{Ptot}
\end{equation}
Substituting them in Eq.~\eqref{DEP} we obtain
\begin{equation}
\mathbf{D} = \varepsilon \mathbf{E} + \mathbf{P}_{\mathrm{dipole}}. \label{DEPunit}
\end{equation}
Now, let us calculate the value of 
$\mathbf{E}$, $\mathbf{P}_{\mathrm{dipole}}$, $\mathbf{D}$,
and the charge density,
considering the unreconstructed AlO$_2$-terminated (001) surface of LaAlO$_3$ as an example,
for the two different unit cell assignments previously discussed.

\begin{figure}[]
\begin{center}
\includegraphics[width=12cm]{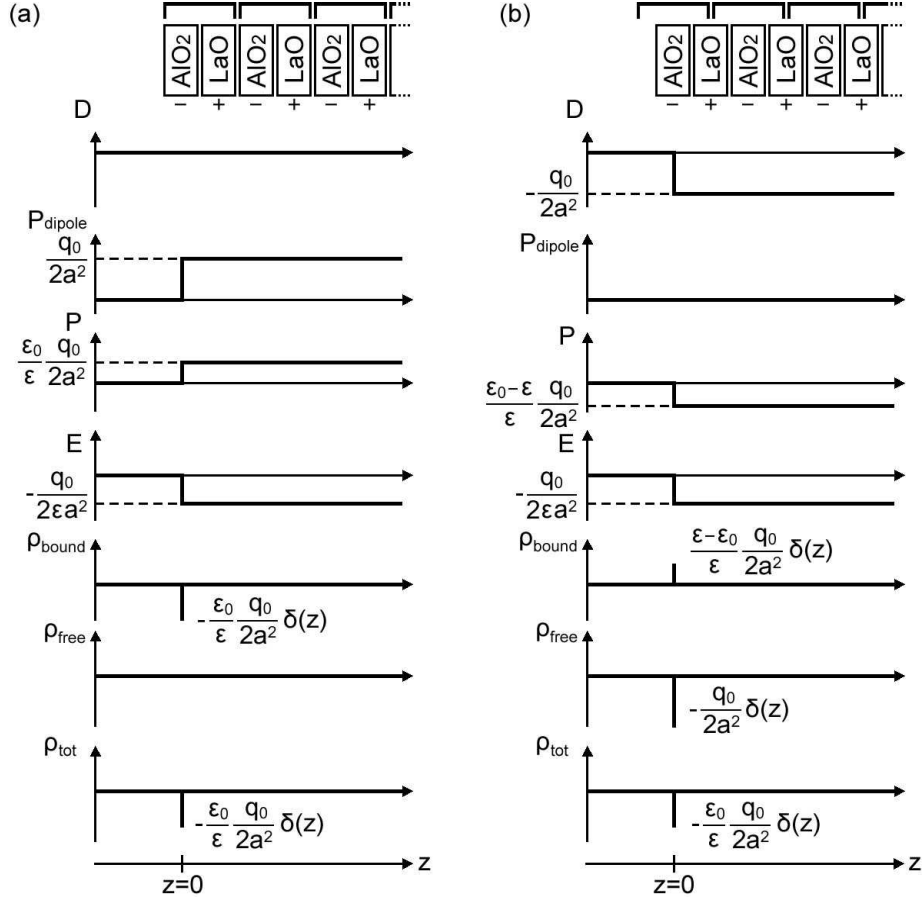}
\caption{\label{local-maxwell} 
Schematic illustration of macroscopic $D$, $P_{\mathrm{dipole}}$, $P$, $E$, $\rho_{\mathrm{bound}}$, $\rho_{\mathrm{free}}$, and $\rho_{\mathrm{tot}}$ close to the (001) surface of LaAlO$_3$.
(a) Taking [(AlO$_2$)$^-$ (LaO)$^+$]  as a unit cell and 
(b) taking [$\frac{1}{2}$(LaO) - (AlO$_2$) - $\frac{1}{2}$(LaO)]  as a unit cell.}
\end{center}
\end{figure}

\subsubsection{Taking [(AlO$_2$)$^-$ (LaO)$^+$]  as a unit cell -- the \textit{polar discontinuity} picture}

The $z$-component of the vectors $\mathbf{D}$, $\mathbf{E}$, and $\mathbf{P}$ are denoted as
$D$, $E$, and $P$.
As shown in Fig.~\ref{local-maxwell}(a), all the ionic charges are included in the unit cells,
and thus bound.
To fix the constant in $E$, the vacuum can be taken as the reference for $E=0$.
Due to the absence of free charges, $D = 0$ from Eq.~\eqref{divD}.
In this case, the [(AlO$_2$)$^-$ (LaO)$^+$] unit cell has a dipole moment of $\displaystyle \frac{q_0 a }{ 2}$, 
where $q_0$ is the elementary charge and $a$ is the pseudocubic lattice constant of LaAlO$_3$.
Considering the unit cell volume $a^3$, 
$P_{\mathrm{dipole}}$ is given by $\displaystyle P_{\mathrm{dipole}} = \frac{q_0}{2a^2}\theta(z) $,
where $\theta(z)$ is the step function.
Equation \eqref{DEPunit} immediately provides $E$ as a function of the position,
namely $\displaystyle E = -\frac{q_0}{2\varepsilon a^2} \theta(z) $.
The total polarization is given by
$ \displaystyle
P = \left( (\varepsilon - \varepsilon _0) \cdot \frac{-q_0}{2\varepsilon a^2} + \frac{q_0}{2a^2}  \right) \theta(z)
= \frac{\varepsilon _0}{\varepsilon} \frac{q_0}{2a^2} \theta(z)  $,
following equations \eqref{Pind} and \eqref{Ptot},
and from Eq. \eqref{divP},
$ \displaystyle
\rho_{\mathrm{bound}}
= - \frac{\mathrm{d}}{\mathrm{d}z} P 
= - \frac{\varepsilon _0}{\varepsilon} \frac{q_0}{2a^2} \delta (z)$,
where $\delta (z)$ is the Dirac $\delta$ function.
This indicates that the system has bound charges of 
$\displaystyle - \frac{\varepsilon _0}{\varepsilon} \frac{q_0}{2a^2}$
at the surface.

\subsubsection{Taking [$\frac{1}{2}$(LaO) - (AlO$_2$) - $\frac{1}{2}$(LaO)]  as a unit cell 
-- the \textit{local charge neutrality} picture}

When taking dipole-free unit cells [Fig.~\ref{local-maxwell}(b)], 
the topmost unit cell is [(AlO$_2$) - $\frac{1}{2}$(LaO)],
which is not a bulk unit cell.
Therefore, the half negative charge which belongs to this unit cell is free, and  $\displaystyle  \rho_{\rm free} = - \frac{q_0}{2a^2} \delta (z)$.
By integrating Eq.~\eqref{divD} and using the boundary condition that $D=0$ in the vacuum,
we obtain $\displaystyle D = - \frac{q_0}{2a^2} \theta (z)$.
Since  $P_{\mathrm{dipole}} = 0$, $E$ is obtained as 
$\displaystyle E = \frac{D - P_{\mathrm{dipole}}}{\varepsilon} = - \frac{q_0}{2\varepsilon a^2} \theta (z)$.
According to Eq.~\eqref{DEP}, the total polarization is
$\displaystyle P = D - \varepsilon_0 E =  \frac{\varepsilon - \varepsilon_0}{\varepsilon}\frac{q_0}{2 a^2}\theta(z)$,
and $\rho_{\rm bound}$ is obtained  by Eq.~\eqref{divP}, namely
$\displaystyle \rho_{\mathrm{bound}} = \frac{\varepsilon - \varepsilon_0}{\varepsilon}\frac{q_0}{2 a^2} \delta(z)$.
This bound charge density appeared as a response to the electric field $E$
arising from $\rho_{\mathrm{free}}$.
Thus we confirm that $\rho_{\rm total}$ is independent of the choice of unit cell.

\section{Further discussions}

\subsection{Effect of interdiffusion}\label{sec-Effect of diffusion}

In real systems, interdiffusion of atoms across the interface is inevitable,
and we should note its effect on the electrostatic stability of the system.
Interdiffusion is a process where atoms are locally exchanged,
and does not change the charge neutrality conditions around the interface,
except for the finite dipoles induced by the modulation of charges.
Since the instability of polar interfaces
can be derived from the lack of charge neutrality around them, 
simple interdiffusion in a finite region cannot compensate it.
That is, local stoichiometric interdiffusion can neither create, nor remove, a potential divergence.
Here, this point is emphasized by considering a simple model.

\begin{figure}[]
\begin{center}
\includegraphics[width=12cm]{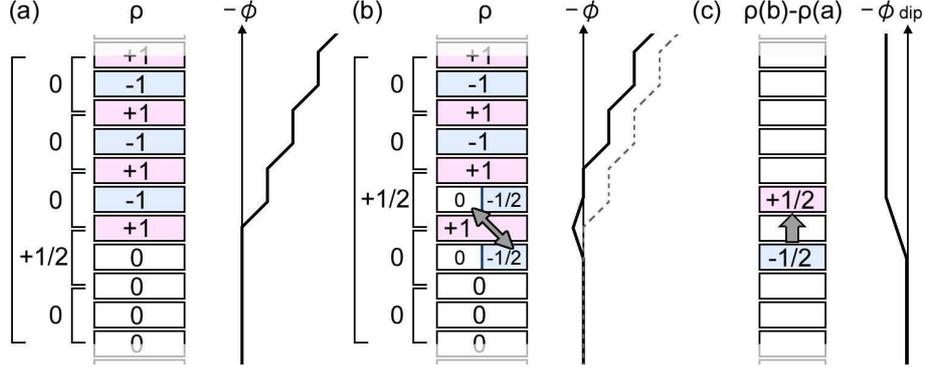}
\caption{\label{local-diffusion} 
Schematic charge structure $\rho$ and electrostatic potential $\phi$ of
(a) an abrupt interface and (b) an interface with interdiffusion.
Five dipole-free unit cells are taken as an example of a cluster covering the interface region.
The dashed line in (b) is the duplication of $\phi$ in (a).
(c) Difference $\rho(\mathrm{b})-\rho(\mathrm{a})$ of the charge distributions in (a) and (b),
and created potential shift $\phi_{\mathrm{dip}}$ by the dipole indicated by the shaded arrow.}
\end{center}
\end{figure}

Figure.~\ref{local-diffusion}(a) shows the charge structure $\rho$ 
and the calculated electrostatic potential $\phi$ of an abrupt polar interface.
By taking dipole-free unit cells, the interface unit cell has a half positive charge per 2D unit cell,
and the system does not keep charge neutrality as it is, and is unstable.
Figure.~\ref{local-diffusion}(b) shows an example of interdiffusion,
where half of the negatively charged layer and half of the charge neutral layer close to the interface
are swapped, compared to the model in Fig.~\ref{local-diffusion}(a).
In this case, the extra positive charge appears at a different position,
but the amount of the charge is conserved, and a similar instability in $\phi$ arises 
as in the abrupt case.

In order to show that interdiffusion does not change the stability or instability of a polar interface, 
it is useful to consider charge neutrality in a cluster
consisting of a sufficiently large number of dipole-free unit cells covering the interface region.
By taking dipole-free unit cells, the electrostatic structure in the bulk can be neglected
to determine the stability of the system.
As an example, let us take a cluster shown in Figs.~\ref{local-diffusion}(a) and (b).
Since the modulation of the charge to achieve the interdiffused model
occurs inside the cluster, the total amount of the charge in the cluster is conserved,
and thus the same instability appears in both cases.

For comparison of these models, let us focus on the difference of the charge distributions
$\rho(\mathrm{b})-\rho(\mathrm{a})$
in the two models, as shown in Fig.~\ref{local-diffusion}(c),
where a dipole with a finite size appears.
Since $\phi$ is calculated by spatially integrating the charge distribution twice,
the difference of $\phi$ in Figs.~\ref{local-diffusion}(a) and (b)
should be equal to the potential shift 
created by the dipole in Fig.~\ref{local-diffusion}(c).
Indeed, $\phi$ in Fig.~\ref{local-diffusion}(b)
has a shift from that in Fig.~\ref{local-diffusion}(a), 
which is the same amount as the dipole shift in Fig.~\ref{local-diffusion}(c).
It should be highlighted that this dipole shift can indeed be an interface-specific additional driving force for interdiffusion, and has been suggested to fundamentally limit the abruptness of some interfaces \cite{Nakagawa}.  
The energy associated with a band offset, for example, can be reduced by forming this dipole.

Finally, we should note the difference between simple interdiffusion and change of the interface composition.
In the interdiffusion process, only exchanging atoms in the finite interface region is allowed,
and therefore the total number of atoms of each element in the region is conserved.
On the other hand, interface composition can be changed, for example by inserting other materials,
segregation of atoms, or creating vacancies. 
For example, the reconstruction model in Fig.~\ref{local-HarrisonGeGaAs}(b)
to compensate the instability of a polar Ge/GaAs interface cannot be achieved by a simple roughening
at the interface: the numbers of Ge, Ga, and As atoms are different compared to 
those in the abrupt model [Fig.~\ref{local-HarrisonGeGaAs}(a)].

In summary, the effect of the interdiffusion
appears as an extra interface dipole moment at the interface from the viewpoint of electrostatics.
However, it does not change the total number of charges at the vicinity of the interface,
and thus cannot screen the charge imbalance at polar interfaces.
In order to compensate the instability of a polar interface, therefore,
introduction of a compositional change or other compensating charge is required.

\subsection{Role of correlation effects}\label{sec-Further discussion}

So far we have tried to simply describe the perovskite polar discontinuity using semiconductor language
by taking dipole-free unit cells and treating the interface unit cell as a $\delta$-dopant.
It should be mentioned that in the presence of strong electron-electron correlations, 
commonly found in transition metal oxides,
it is formally impossible to draw semiconductor energy band diagrams 
based on the single-particle picture of independent electrons\footnote{
Owing to the relatively small (still non-negligible) correlation effects 
in SrTiO$_3$ with the Ti $3d^0$ configuration and weak 2$p$-3$d$ hybridization in the coherent state of doped SrTiO$_3$ \cite{Ishida2008},
the schematic band diagram shown in Fig.~\ref{local-delta-LAOSTO} is still reasonably valid,
approximating SrTiO$_3$ to be a band semiconductor.} \cite{Eskes1991, Fujimori1992}.
Also, in order to draw band diagrams,
we should know which part of the charges are to be assigned as free carriers,
which is non-trivial in correlated systems, such as Mott insulators.
For example, a perovskite with 1 electron per unit cell can give rise to an effective carrier density ranging from $\sim1\times10^{22}{\rm ~cm}^{-3}$ to zero, depnding on the correlation strength.

However, the \textit{local charge neutrality} picture can still provide important information 
on the interface charge structure of transition metal oxides,
since correlations cannot change the amount of total charge.
While the distribution of this charge may be significantly modified by correlation features in the electronic compressibility \cite{Oka2005},
the charge can still be determined in a cluster consisting of
a sufficiently large number of dipole-free unit cells.
This is because the material outside of the charge modulation region
consists of dipole-free unit cells of bulk, which are charge neutral and thus stable.

\subsection{Quadrupolar discontinuity}
\label{sec-quadrupolar discontinuity}

Thus far, we considered the stability of polar discontinuities,
and showed that the amount of charge needed at the interface can be determined
by taking dipole-free unit cells and considering local charge neutrality in each unit cell.
Of course, local charge modulation is allowed once neutrality is obtained,
since it does not violate global charge neutrality.
This dipole energy arising from the modulation of charges
determines the real charge distribution in the system.

\begin{figure}[]
\begin{center}
\includegraphics[width=12cm]{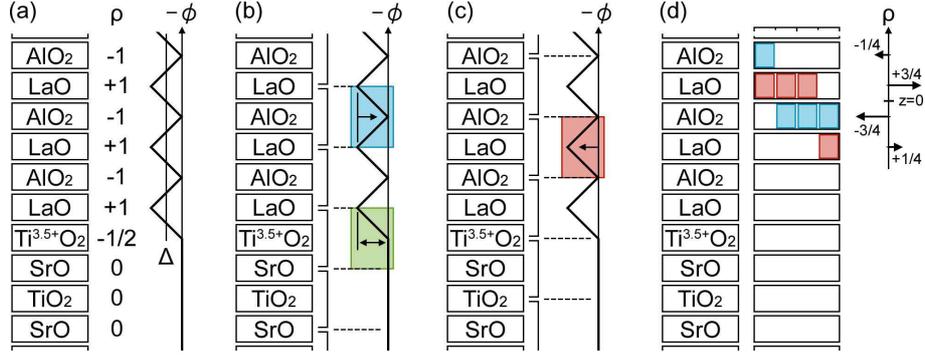}
\caption{\label{Appendix-quadra} 
(a) Schematic structure, charge density $\rho$, and electrostatic potential $\phi$ 
of a LaAlO$_3$/SrTiO$_3$ interface with a Ti$^{3.5+}$O$_2$ interface layer,
showing a finite shift $\Delta$.
Taking (b) [($\frac{1}{2}$LaO)$^{+1/2}$-AlO$_2^{-1}$-($\frac{1}{2}$LaO)$^{+1/2}$] 
and (c) [($\frac{1}{2}$AlO$_2$)$^{-1/2}$-LaO$^{+1}$-($\frac{1}{2}$AlO$_2$)$^{-1/2}$] as a unit cell.
The blue and pink shaded areas in (b) and (c) show the unit cells with opposite signs of quadrupole moment, respectively,
and the green shaded area in (b) shows the unit cell with a finite dipole moment.
(d) An example of dipole- and quadrupole-free unit cells.}  
\end{center}
\end{figure}

Here, we note that there is an intrinsic dipole shift at the interface 
arising from the charge stacking sequence,
which should be distinguished from this charge modulation.
For example, consider a LaAlO$_3$/SrTiO$_3$ (001) interface with Ti$^{3.5+}$ at the interface
to solve the instability of the polar interface, as shown in Fig.~\ref{Appendix-quadra}(a).
Although there is no potential divergence,
a finite shift $\Delta$ remains between the two materials
when considering the averaged electrostatic potential on both sides.

The origin of this potential shift 
can be understood based on the discontinuity of the quadrupole moment
of the unit cells.
In one dimension, the quadrupole moment density $Q$ is defined as
$\frac{\mathrm{d}}{\mathrm{d} x} Q =  P$, where $P$ is the dipole moment density. 
It has the same form as that of Gauss' law connecting the dipole moment and the charge
--- i.e., the source of the quadrupole moment is the dipole moment.
Therefore, following the same argument as in Section~\ref{sec-Unit cells in ionic crystals},
a finite interface dipole moment appears 
at a quadrupolar discontinuity, when the quadrupole moment is different on the two sides.

Moreover, this quadrupole moment is proportional to the potential shift at that point in the absence of
free charge, since $D=0$ and thus $\varepsilon_0 E = -P$.
Therefore if the unit cell has a finite quadrupole moment,
that means the averaged potential in the unit cell is shifted by the corresponding value.
Note the quadrupole moment does not induce a potential shift outside of the unit cell,
while the dipole moment does change.

For example, [($\frac{1}{2}$LaO)$^{+1/2}$-AlO$_2^{-1}$-($\frac{1}{2}$LaO)$^{+1/2}$],
a dipole-free unit cell of LaAlO$_3$ (001) [Fig.~\ref{Appendix-quadra}(b)],
has a finite quadrupole moment,
while the charge neutral stacking of SrTiO$_3$
creates no quadrupole moment in the unit cells\footnote{Absence of covalency is assumed.}.
This difference of quadrupole moment between the two sides
adds a finite potential shift to the averaged potential in the unit cell.
Figures~\ref{Appendix-quadra}(b) and (c) show two ways of taking dipole-free unit cells
in cubic perovskites,
where the sign of the quadrupole moment is opposite.
The choice of unit cells cannot change the electrostatic potential,
and indeed this difference of the quadrupole moment of the unit cells 
is compensated by the dipole moment of the interface unit cell in Fig.~\ref{Appendix-quadra}(b).

The ambiguity of the quadrupole moment in the unit cells is reminiscent 
of the uncertainty of the dipole moment, as discussed in Section~\ref{sec-Unit cells in ionic crystals},
and we can take the same prescription as we did previously --- taking quadrupole-free unit cells.
It is more complicated since these quadrupole-free unit cells 
should not have a dipole moment at the same time,
in order to avoid the instability of the polar discontinuity,
and Fig.~\ref{Appendix-quadra}(d) shows one way to take such unit cells.
Note it is always possible to take unit cells which do not have 
either dipole or quadrupole moments in one direction
in any bulk crystal\footnote{This can be easily proved from the periodicity of the charge structure and 
the macroscopic charge neutrality of the lattice by using the intermediate-value theorem.}.
Once we take this dipole- and quadrupole-free unit cell, the size of the interface dipoles 
can be determined locally by considering the dipole moment in the interface unit cell(s),
since there is no shift in the averaged potential in the bulk unit cells due to the absence of  
quadrupole moment.

\section{Summary}

As pointed out by Tasker, the order of the ionic charge stacking
plays an important role for the stability of the surfaces and interfaces,
and sometimes they show a diverging potential when following bulk compositions and electronic states.
In semiconductors where the number of ionic charges are fixed,
they are usually stabilized by atomic reconstructions, as shown by Harrison \textit{et al.}.
However, in transition metal oxides, electronic reconstructions provide
another possibility to reconcile the instability.
Exploiting this degree of freedom has been a central topic of recent research in oxide heterostructures, but with much debate over the relative contribution of this effect, as compared to those common to all interfaces such as defects and diffusion.

We hope the reader will agree that the attribution of various experimental measurements in real materials to purely the polar discontinuity, 
as opposed to chemical defects, is not really a valid separation. 
It is rather governed by the response of the total system to the electro-chemical potential (in the chemistry sense), subject to specific thermodynamic, kinetic, and geometric boundary conditions.
Given the freedom to define the unit cell, taking dipole-free units provides a convenient format for this perspective. 
While not explicitly discussed here, interface screening by local lattice polarization can be directly incorporated, including subtleties that arise at very short length scales.  
Thus these topics connect to current research in surface and interface phenomena for ferroelectrics and multi-ferroics.  
Furthermore, recent discussion of the role of film stoichiometry can be understood together with these other aspects  \cite{Chambers2010317}. 
Ultimately, the requirement to stabilize polar discontinuities is independent of
whether the charge is free or bound, itinerant or localized, and electronic or ionic.
Perhaps the central message of this simple review chapter is that one cannot claim atomically precise oxide heterostuctures without understanding how the electrostatic boundary conditions are accomodated. This is both a problem to be solved, as well as a fascinating synthetic opportunity.

\vspace{30pt}
\noindent {\bf Acknowledgments:} We thank our many colleagues and collaborators in this field who have helped develop these topics, and T. Yajima for pointing out the quadrupole-dipole connection. We acknowledge support from the Japan Science and Technology Agency, the Japan Society for the Promotion of Science, and the Department of Energy, Office of Basic Energy Sciences, under contract DE-AC02-76SF00515 (H.Y.H.).

\newpage


\end{document}